\documentclass[amsmath,amssymb,preprint,aps]{revtex4}
\usepackage{epsfig,graphics,graphicx,amsmath}

\begin{document}

\title{On the effect of heterogeneity in stochastic interacting-particle systems}
\author{Luis F. Lafuerza$^\ast$ and Raul Toral}

\affiliation{Instituto de F{\'\i}sica Interdisciplinar y Sistemas Complejos (IFISC), CSIC-UIB,  Campus UIB, E-07122 Palma de Mallorca, Spain  }

\date{\today}

%89.65.Gh  :Economics; econophysics, financial markets, business and management
%05.40-a :Fluctuation phenomena, random processes, noise, and Brownian motion   89.75.-k : complex systems;     05.10.Gg : Stochastic analysis methods (Fokker-Planck, Langevin, etc.)

\begin{abstract}
We study stochastic particle systems made up of heterogeneous units. We introduce a general framework suitable to analytically study this kind of systems and apply it to two particular models of interest in economy and epidemiology. We show that particle heterogeneity can enhance or decrease the size of the collective fluctuations depending on the system, and that it is possible to infer the degree and the form of the heterogeneity distribution in the system by measuring only global variables and their fluctuations. Our work shows that, in some cases, heterogeneity among the units composing a systems can be fully taken into account without losing analytical tractability
 \end{abstract}

\maketitle

Most real systems are made up of heterogeneous units. Whether considering a population of cells \cite{diversitycells,diversitycellsrev}, a group of people \cite{granovetter,kirmanrepagent} or an array of lasers \cite{nonidlasers,nonidlasersstrogatz} (to name just a few examples), one never finds two units which behave exactly in the same way. Despite this general fact, quantitative modeling most often assumes identical units, since this condition seems necessary for having analytically tractable models. Moreover, in the general framework of complexity science, systems very often can be modeled only at a stochastic level, since a complete knowledge of all the variables, the precise dynamics of the units and the interaction with the environment is not available. One way to include system heterogeneity is to consider that the interactions between the units are not homogeneous but mediated by some complex network, an approach that has attracted enormous attention in the last years \cite{barabasirev,yamirnetworks}. An issue that has been less studied, beyond the role of particle heterogeneity in deterministic systems \cite{divresonance,quenchednoise,pyoung,epidemic},  is the heterogeneity in the behavior of the particles themselves in stochastic models. Some exceptions include the recent reference \cite{redner}, where the authors analyze the effect of heterogeneous transition rates on consensus times in the voter model, and works considering the effect of a few ``committed'' individuals in this and related models \cite{mobilia,minor}. In the context of statistical physics, the combined effects of stochasticity and heterogeneity have been considered, for example, in random-field Ising models and spin glasses \cite{fisher,Young,Parisi} or in diffusion in disordered media \cite{bouchaud90,avraham-havling00}. We aim here at developing a general framework for the analytical study of stochastic systems made up of heterogeneous units, applicable beyond equilibrium models or Hamiltonian systems and suitable for a general class of complex systems of recent interest and at identifying some generic effects of  particle heterogeneity on the macroscopic fluctuations.

In this work we will show that the combined effect of stochasticity and heterogeneity can give rise to unexpected, non-trivial, results. While, based on na\"ive arguments, one should conclude that global fluctuations increase in heterogeneous systems, we will show that in some systems of stochastic interacting particles fluctuations actually decrease with the degree of heterogeneity. Moreover, we will see that it is possible to infer the degree of particle heterogeneity (or ``diversity'') by measuring only global variables. This is an issue of great interest when one has access only to  information at the macroscopic, population level, since it allows one to determine if heterogeneity is a relevant ingredient that needs to be included in the modeling. In this way, heterogeneity can be included when its presence is implied by the data and it does not enter as an extra free parameter.  We will study first the simple case of independent particles; then we will consider the general case of interacting particles and develop an approximated method of general validity to analytically study these systems; next, as a way of example, this method will be applied to two particular models of interest in economy and epidemiology.

Our starting point is a stochastic description of a system composed by $N$ non-identical units, which we call generically ``particles'' or ``agents". Each particle is characterized by a constant parameter $\lambda_i$ ($i=1,\dots, N$); the value of this parameter differs among the particles and it is the source of heterogeneity considered. Although there are more general ways of including heterogeneity, we will stick to this type of parametric heterogeneity \cite{parametric_heterogeneity} because it is simple yet rather general. For simplicity, we assume that each particle can be in one of two possible states and define $s_i(t)=0,1$ as the variable describing the state of particle $i$ at time $t$ (the two-states assumption will be relaxed later). The collective state of the system is given by the total number $n(t)=\sum_{i=1}^N s_i(t)$ of particles in state $1$. Sometimes, one does not have access to the individual dynamics and can only access experimentally the value of $n(t)$.
We are interested in the statistical properties of this global variable and how do they depend on the degree of heterogeneity in the system. We will often refer to $n(t)$ as the {\sl macroscopic} variable and to the $s_i(t)$'s as the {\sl microscopic} ones.
\section{Results}
\subsection{Independent Particles}\label{section:independent}
\begin{figure}
\centering
\includegraphics[scale=0.4,angle=0,clip]{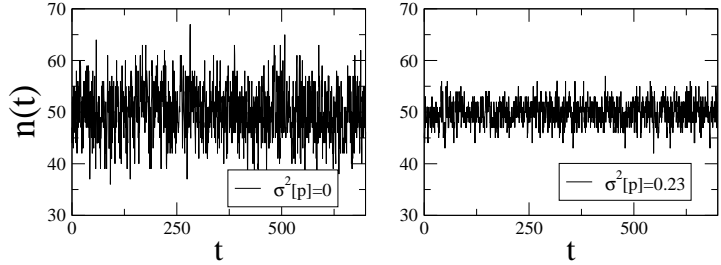}
\caption{Time series for the global variable $n(t)$ of a system of identical (left panel) and heterogeneous (right panel) particles, for a system of $N=100$ particles. The parameters were set as $r^+_i=1, r^-_i=1/p_i-1$, with $p_i=1/2$ in the case of identical particles (left panel) and $p_i$ chosen from a symmetric Beta distribution $f(p)=\frac{\Gamma(\alpha)^2}{\Gamma(2\alpha)}[p(1-p)]^{\alpha-1}$, with $\alpha=0.05$, being the sample mean and variance equal to $\overline{p}=0.501$, $\sigma^2[p]=0.23$, respectively. Note that the fluctuations of the average state are larger in the case of identical particles.
 \label{traject}} 
\end{figure}

We study first the case in which  particles jump independently from state $0$ to $1$ and vice-versa, schematically:
\begin{eqnarray}
0 {{r_i^+ \atop \longrightarrow}\atop{}} 1, \hspace{0.5cm} 1{{r_i^- \atop \longrightarrow}\atop{}} 0,
\end{eqnarray}
with rates that depend on the value of the heterogeneity parameter, $r_i^{\pm}= r^{\pm}(\lambda_i)$. The probability $p_i(t)$ for particle $i$ to be in state $1$ at time $t$ obeys the linear rate equation $\displaystyle \frac{d p_i}{dt}=-r_i^-p_i+r_i^+(1-p_i)$. In the case of constant rates, the solution is: $p_i(t)=\frac{r_i^+}{r_i}(1-e^{-r_it})+p_i(0)e^{-r_it}\label{pi}$, with $r_i\equiv r_i^++r_i^-$. The results derived below apply equally if the rates depend on time or on the time that the particle has been in its current state (if the rate depends on the time $a$ that the particle has been on its current state, the steady-state probability of finding the particle at state $1$ is $p_{i,\textrm{st}}=\frac{\Lambda^-_i}{\Lambda^+_i+\Lambda^-_i}$ with $\Lambda^\pm_i=\int_0^\infty dt\,e^{-\int_0^tda\,r_i^\pm(a)}$) . Using particle independence and that the moments with respect to realizations of the stochastic process of the random variable $s_i$ are given by $\langle s_i^k\rangle=1^kp_i+0^k(1-p_i)=p_i$, one obtains that the average and variance of the global variable $n$ are: 
\begin{eqnarray}
\langle n(t)\rangle&=&\sum_{i=1}^N p_i(t)= N\overline{p(t)},\label{medparticular}\\
\sigma^2[n(t)]&=&\sum_{i=1}^N \left(p_i(t)-p_i(t)^2\right)= N\left(\overline{p(t)}-\overline{p(t)^2}\right),\label{varparticular}
\end{eqnarray}
where the overline denotes an average over the population, $\overline{g}\equiv\frac{1}{N}\sum g_i$.
If we consider a system where all particles are identical (i.e. have the same values for the internal parameter $\lambda_i=\lambda_j,\forall i,j$), and keep the same average value $\langle n(t)\rangle$ for the global variable at time $t$, the variance would be $\sigma_\text{id}^2[n(t)]=N\overline{p(t)}\left(1-\overline{p(t)}\right)\ge \sigma^2[n(t)]$. We obtain the somehow counterintuitive result that a system of heterogeneous independent particles displays smaller fluctuations in its collective variable than another system with identical particles. This effect is illustrated in figure~\ref{traject}. The reduction in the variance of the collective variable is $N$ times the variance of $p_i$ over the population:
\begin{equation}
 \sigma_\text{id}^2[n(t)]-\sigma^2[n(t)]=N\left(\overline{p(t)^2}-\overline{p(t)}^2\right),
\end{equation}
which is of the same order, $O(N)$, as the variance itself, giving a non-negligible correction. 

Reading the previous formula backwards, one realizes that the moments of the collective variable give information about the degree of heterogeneity in the system:
\begin{equation}
\overline{p(t)^2}-\overline{p(t)}^2=\frac{\langle n(t)\rangle-\langle n(t)\rangle^2/N-\sigma^2[n(t)]}{N}.
\end{equation}
This expression is general, regardless the specific form in which $p_i$ is distributed over the population. Higher moments of the heterogeneity distribution are also related to higher moments of the collective variable. This allows to infer the skewness, kurtosis and higher order characteristics of the heterogeneity distribution by measuring only global variables and their fluctuations. In the Supplementary Information it is shown that an equivalent result is obtained generically for $k$-state systems for $k>2$.

Besides the moments, one can derive the full probability distribution of the global variable. The generating function for the one particle probability distribution is $g_i(z)=\sum_{s_i=0}^1z^{s_i}P(s_i)=1-p_i+p_iz$ and the generating function $G(z)=\sum_{n=0}^Np(n)z^n$ of the sum of independent random variables is the product of their generating functions, $
 G(z)=\prod_{i=1}^Ng_i(z)
 $.
Expanding in powers of $z$ we can obtain the probability distribution for $n$: $P(n)=\frac{1}{n!(N-n)!}\sum_{{\bf i}\in S_N}\prod_{\ell=1}^np_{i_\ell}\prod_{\ell=n+1}^N(1-p_{i_\ell})$, being ${\bf i}=(i_1,\dots,i_N)$ and $S_N$ the group of permutations of $N$ elements.

The model studied in this subsection, despite its simplicity, offers a reduced description of generic systems of non-interacting multi-stable units subject to fluctuations. The results obtained here are directly relevant if one is interested in the collective properties of one such system when the units are non-identical.  One can reasonably argue that the independence property is too unrealistic for the study to be of any practical interest. We will be considering more complicated cases including non-independent units in the rest of the paper. However, this simple model presents in isolation a mechanism, spontaneous transitions, that can play a role in more complicated and relevant systems (we will see this later). The simplicity of the model allows us to understand the effect of heterogeneity in this mechanism, and will give us insight in the role of heterogeneity in the behavior of more complicated systems.

\subsection{Two types of uncertainties}
We will now discuss the situation in which the particular values of the parameter of each particle are not known. This introduces an additional  source of uncertainty. For simplicity, we will focus on the $2$-states independent particles system considered before, but the discussion applies as well to general systems of interacting particles. This discussion will also allow us to take a closer look at the results obtained and clarify their meaning and relevance in different settings.

Often, one does not know the value of the parameter $\lambda_i$ of each individual particle, but has some idea about how this parameter is distributed on the population, perhaps its probability distribution (obtained for example by measuring individual behavior in an equivalent system). Here, we will assume that the $\lambda_i$'s are independent and identically distributed random variables with a given probability density $f(\lambda)$. In this case, $\langle n\rangle$ and $\sigma^2[n]$ are themselves random variables that, as shown above, depend on the particular realization of the $\lambda_i$'s. The expected values of these quantities are obtained by averaging Eqs.(\ref{medparticular},\ref{varparticular}) over the distribution of the individual parameters:
\begin{equation}
\widehat{\langle n(t)\rangle}=N\widehat{p(t)},\,\,\,\,
\widehat{\sigma^2[n(t)]}=N\left(\widehat{p(t)}-\widehat{p(t)^2}\right)\label{varmed},
\end{equation}
where the hat denotes an average with respect to $f(\lambda)$, $\widehat{g}\equiv\int d\lambda g(\lambda)f(\lambda)$. Again the variance is smaller than for a system of identical particles with the same mean value, namely, $\sigma^2_\text{id}[n(t)]-\widehat{\sigma^2[n(t)]}=N\left(\widehat{p(t)^2}-\widehat{p(t)}^2\right)$. 

If we average the probability $P(n)$ over the distribution of parameters we obtain a simple form for the probability of the global variable $n$:
\begin{eqnarray}
 &&\widehat{P(n)}=\int d\lambda_1\dots d\lambda_NP(n|\lambda_1,\dots,\lambda_N)f(\lambda_1)\dots f(\lambda_N)\nonumber\\
&=&\binom{N}{n}\widehat{p\,}^n\left(1-\widehat{p\,}\right)^{N-n}\label{averagedistribindep},
\end{eqnarray}
a binomial distribution with parameter the average $\widehat{p}=\int d\lambda p(\lambda)f(\lambda)$ over the distribution $f$. The variance of this distribution is
\begin{equation}
 \sigma^2[n(t)]_\text{tot}=N\left(\widehat{p(t)}-\widehat{p(t)}^2\right)\label{vartotal},
\end{equation}
equal to the variance one would obtain in a system of identical particles with the same average, $N\widehat p(t)$, a result in apparent contradiction with (\ref{varmed}). However, we should note that they refer to different things: Expression (\ref{varmed}) gives the average variance when the parameter values are given, so measuring the average uncertainty in $n$ due to the stochastic nature of the process.  (\ref{vartotal}), in addition to the uncertainty coming from the stochasticity of the process, also includes the uncertainty on the parameter values.

The two expressions are related by the law of total variance:
\begin{eqnarray}
 \sigma^2[n]_{tot}&=&\widehat{\sigma^2[n]}+\sigma^2[\langle n|\lambda_1,\dots,\lambda_N\rangle]\nonumber\\
 &=&N(\widehat{p}-\widehat{p^2})+\sigma^2\left[\sum_i p_i\right]
=N(\widehat{p}-\widehat{p}^2).
\end{eqnarray}
In $\sigma^2[\langle n|\lambda_1,\dots,\lambda_N\rangle]$, the variances are taken over the distribution of the $\lambda_i$'s.
If we are considering a particular system, the temporal fluctuations (all the systems considered in this paper are ergodic, so we can think on averages over time or over the realization of the stochastic process interchangeably) in $n$ will come only from the intrinsic stochasticity, and expressions (\ref{varparticular},\ref{varmed}) are the ones that measure it. Expressions (\ref{averagedistribindep},\ref{vartotal}) are appropriate only if we are considering an ensemble of systems with a distribution of parameters and our different measurements may come from different systems in the ensemble.

\subsection{Formulation of the general method}
Let us now consider a general system of interacting heterogeneous particles. The stochastic description now starts from a master equation for the $N$-particle probability distribution:
\begin{eqnarray}
&&\frac{d P(s_1,\dots,s_N)}{dt}=\sum_{i=1}^{N}(E_i-1)\left[s_i r_i^{-}P(s_1,\dots,s_N)\right]\nonumber\\
&&+\sum_{i=1}^{N}(E_i^{-1}-1)\left[(1-s_i)r_i^{+}P(s_1,\dots,s_N)\right]\label{mastereqgen},
\end{eqnarray}
with step operators defined as $E_i^kF(s_1,...,s_i,...s_N)=F(s_1,...,s_i+k,...,s_N)$. The transition rates $r_i^{\pm}$ might now depend on the state of any other particle (this is how interactions enter in the model). From  Eq.(\ref{mastereqgen}) one can derive for the moments and correlations:
\begin{eqnarray}
 \frac{d\langle s_i\rangle}{dt}&=&\langle r_i^{+}\rangle-\langle (r_i^{-}+r_i^{+})s_i\rangle\label{momentsgeneral}\\
\frac{d\langle s_is_j\rangle}{dt}&=&-\langle q_{ij}s_js_i\rangle+\langle r_i^{+}s_j\rangle+\langle r_j^{+}s_i\rangle%+\delta_{i,j}\left[\langle s_ir_i^-\rangle+\langle(1-s_i)r_i^+\rangle\right].
\label{correlationsgeneral}
\end{eqnarray}
with $q_{ij}=r_i^{-}+r_j^{-}+r_i^{+}+r_j^{+}$ and $i\ne j$ in the second equation (recall that $s_i^2=s_i$). 
In general, if the transition rates depend on the state variables $s_i$, these equations are not closed since they involve higher order moments, and some approximation method is needed to proceed. Systematic expansions in $1/N$, including van Kampen's $\Omega$-expansion \cite{VK}, are not applicable, since variables $s_i=0,1$ are not extensive. In the following, we introduce an approximation suitable for the analytical treatment of systems of globally coupled heterogeneous particles.

We assume that the $m$-particle correlations $\sigma_{j_1,...,j_m}(t)=\langle \delta_{j_1}(t)\cdots\delta_{j_m}(t)\rangle$ with

$\delta_j(t)= s_j(t)-\langle s_{j}(t)\rangle$
 scale with system size as 
\begin{equation}
 \sigma_{j_1,...,j_m}(t)=O(N^{-m/2}),\hspace{0.3cm} \text{for } j_k\neq j_l.\label{ansatz}
\end{equation}
Using this ansatz one can close the system of equations (\ref{momentsgeneral},\ref{correlationsgeneral}) for the mean values and the correlations. This is shown in the Supplementary Information for general transition rates of the form $f(s_1/N,\dots,s_N/N)$. 
 
While the resulting equations for the average values $\langle s_i(t)\rangle$ coincide with the mean-field rate equations usually formulated in a phenomenological way\cite{Young,epidemic}, our formulation allows us  to compute the correlations and include, if needed, higher order corrections in a systematic way.

Assumption (\ref{ansatz}) can be justified noting that it is consistent with  $\sum_{j_1,...,j_m}\sigma_{j_1,...,j_m}=\langle(n-\langle n\rangle)^m\rangle=O(N^{m/2})$ which follows from van Kampen's splitting of the global variable $n=N \phi+N^{1/2}\xi$, with $\phi$ deterministic and $\xi$ stochastic. Details are given in the Supplementary Information. The global variable $n$ is extensive and it is expected to follow van Kampen's ansatz in many cases of interest. Note, however, that since there is not a closed description for the macroscopic variable $n$, one can not use van Kampen's expansion, and our approach extends the implications of this splitting of the macroscopic variable to the correlations of the microscopic state variables. For simplicity, we have focused on $2$-states systems and assumed a constant number of particles. Systems with $k$ states are also expected to follow ansatz (\ref{ansatz}), since the scaling of the global variable is not limited to $2$-sates systems. The case of variable, but bounded, number of particles can be included straightforwardly by considering an extra state. The unbounded case can also be considered performing an appropriate limit. If the system has some spatial structure, the ansatz (\ref{ansatz}) is not expected to be valid, and some decay of the correlations with the distance is expected instead; this interesting situation is left for future work.

We will proceed by applying the presented method to analyze the role of heterogeneity in two models previously considered in the literature that apply to contexts in which the assumption of identical agents can hardly be justified: stock markets and disease spreading. We will focus on the steady-state properties of both models, skipping transient dynamics.
\subsection{Application to Kirman Model}
Kirman's model \cite{kirman} was proposed to study herding behavior in the context of stock markets and collective dynamics on ant colonies. In the stock market context, agent $i$ can be in two possible states (e.g. $0\equiv$``pessimistic'' -with regard to future market price- and $1\equiv$``optimistic'') and it can switch from one to the other through two mechanisms: spontaneous transitions at a rate $\epsilon$, and induced transitions at a rate $N^{-1}\sum_j\lambda_j(1-\delta_{s_i,s_j})$, being $\lambda_j$ the ``influence" of agent $j$ on other agents. The case $\epsilon=0$ corresponds to the voter model\cite{votermodel}. 
In the original formulation, all agents have the same influence, i.e. $\lambda_i=\lambda_j, \forall i,j$. We generalize the model allowing the parameter $\lambda_i$ to vary between agents. In \cite{alfarano}, the effect of heterogeneity was explored numerically, but not in a systematic way.

This model is interesting for us because it incorporates in a simple way two basic processes: spontaneous transitions and induced transitions. As we will see, due to its simplicity, a full analytical treatment is possible that will, in turn, allow us to obtain a deeper insight into the general effect of heterogeneity in systems of interacting particles. 

\begin{figure}
\centering
\includegraphics[scale=0.35,angle=0,clip]{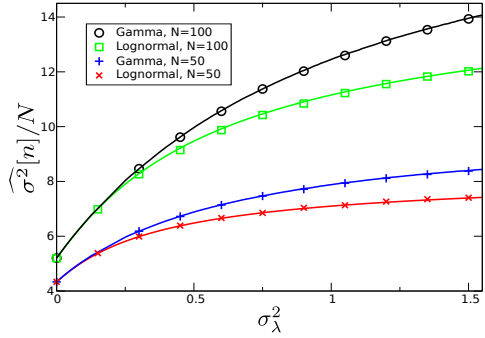}
\caption{Variance of the number of agents in state $1$ as a function of the variance of the influence  parameter $\lambda$ in Kirman's model with distributed influence. Numerical simulations (symbols) and theoretical analysis (lines), Eq.(\ref{kirmanexact}), for different number of agents $N$ and $\epsilon=0.01$. $\lambda_i$ are independent random variables  distributed according to a lognormal or a gamma distribution with mean $\overline{\lambda}=0.5$ and variance $\sigma^2_\lambda$. The results  have been averaged over $2\times10^4$ for $N=50$ and $10^4$ for $N=100$ realizations of the distribution of parameters.
 \label{kirmanaverage}} 
\end{figure}

The master equation for the process is of the form (\ref{mastereqgen}), with rates given by:
\begin{equation}
 r_i^{+}=\epsilon+N^{-1}\sum_k\lambda_ks_k,\,\,r_i^{-}=\epsilon+N^{-1}\sum_k\lambda_k(1-s_k)
\end{equation}
From (\ref{momentsgeneral}) the averages and correlations obey:
\begin{eqnarray}
 \frac{d\langle s_i\rangle}{dt}&=&\epsilon-(2\epsilon+\overline{\lambda})\langle s_i\rangle+N^{-1}\sum_{k}\lambda_k\langle s_k\rangle\label{Kirmanaverage},\\
\frac{d\sigma_{i,j}}{dt}&=&-2(2\epsilon+\overline{\lambda})\sigma_{i,j}+N^{-1}\sum_k\lambda_k\left(\sigma_{i,k}+\sigma_{j,k}\right)\label{Kirmancorrelations}
\end{eqnarray}
for $i\ne j$ and $\sigma_{i,i}=\langle s_i\rangle (1-\langle s_i\rangle)$. 
Note that, due to the particular form of the rates, these equations do not involve higher-order moments. This is a simplifying feature of this model that allows one to obtain exact expressions. The first equation leads to a steady state value $\langle n\rangle_\textrm{st}=\frac{N}{2}$ (a property that comes from the symmetry $0\leftrightarrow1$). Using the relation $\sigma^2[n]=\sum_{i,j}\sigma_{i,j}$ we obtain (see Supplementary Information) that the variance in the steady state is:
\begin{equation}
 \sigma^2_\text{st}[n]=\frac{N}{4}\left[1 + \frac{2\overline{\lambda}(1-N^{-1})}{4\epsilon+\overline{\lambda}} +\frac{ (N-3+2N^{-1})\overline{A}}{2\epsilon+\overline{A}}\right]\label{kirmanexact}
\end{equation}
with $A_i=\frac{\lambda_i^2}{N(4\epsilon+\overline{\lambda})+2\lambda_i}$. The leading-order term,
$
 \sigma^2_\textrm{st}[n]=\frac{N}{4}\left[1+\frac{\overline{\lambda}}{2\epsilon}+\frac{\sigma^2[\lambda]}{2\epsilon\left(4\epsilon+\overline{\lambda}\right)}\right]+O(N^0)
$,
with $\sigma^2[\lambda]=\overline {\lambda^2}-\overline\lambda^2$, can also be readily obtained using the ansatz (\ref{ansatz}). Note that the presence of heterogeneity increases collective fluctuations. In Fig.\ref{kirmanaverage} we compare expression (\ref{kirmanexact}) with results coming from numerical simulations. 

\begin{figure}
\centering
\includegraphics[scale=0.25,angle=0,clip]{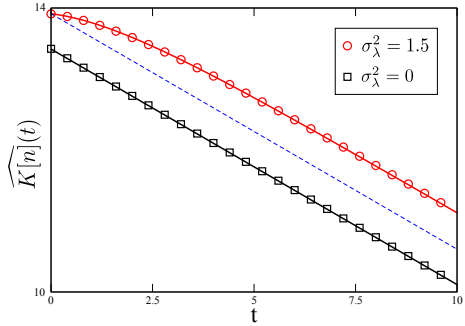}
\caption{Correlation function (in log-linear scale) for Kirman's model with distributed influence. Results coming from numerical simulations (symbols) and theory (Eq.(\ref{kirmancorrfunc}), solid lines). Note that when heterogeneity is present ($\sigma^2_\lambda=1.5$) the correlation function departs from purely exponential decay (displayed as a dashed line). Data for $\sigma^2_\lambda=0$ have been moved up $5.5$ units vertically for better visualization. Parameters values are $\epsilon=0.01$, $N=100$. $\lambda_i$ are independent random variable distributed according to a gamma with mean $\widehat{\lambda}=0.5$ and variance, $\sigma^2_\lambda$, indicated in the figure. A simple fit of expression (\ref{kirmancorrfunc}) to the $\sigma^2_{\lambda}=1.5$ data gives $\overline{\lambda}=0.50, \epsilon=0.0099$.
 \label{kirmancorrf}} 
\end{figure}

In this case, the knowledge of $\langle n\rangle_\text{st}$ and $\sigma^2_\text{st}[n]$ alone does not allow to infer the degree of heterogeneity present in the system, unless one knows from other sources $\overline{\lambda}$ and $\epsilon$ and it is not possible to conclude whether the observed fluctuations have a contribution due to the heterogeneity of the agents. However, the steady-state correlation function $K[n](t)\equiv\langle n(t)n(0)\rangle_\textrm{st}-\langle n\rangle_\textrm{st}^2$, does include a term that allows to infer the possible heterogeneity. $K[n](t)$ is obtained integrating Eq.(\ref{Kirmanaverage}) and performing the appropriate conditional averages (see Supplementary Information):  
\begin{equation}
 K[n](t)=\left(\sigma^2_\text{st}[n]-u\right)e^{-(2\epsilon+\overline{\lambda})t}+ue^{-2\epsilon t},
 \label{kirmancorrfunc}
\end{equation}
with $u\equiv\frac{2\epsilon+\overline{\lambda}}{\overline{\lambda}(1-1/N)}(\sigma^2_\text{st}[n]-N/4)$. The departure from a pure exponential decay signals the presence of heterogeneity (for identical particles $u=\sigma^2_\text{st}[n]$). Fitting this expression to data one can obtain $\sigma^2_\text{st}[n]$ and the parameters $\epsilon,\,\overline\lambda$. Then, the use of expression (\ref{kirmanexact}) would yield $\sigma^2[\lambda]$. In Fig.\ref{kirmancorrf} we show that the numerical simulations indeed support the existence of two exponential decays for the correlation function.

\subsubsection{Other ways to introduce heterogeneity}
\begin{figure}
\centering
\includegraphics[scale=0.25,angle=0,clip]{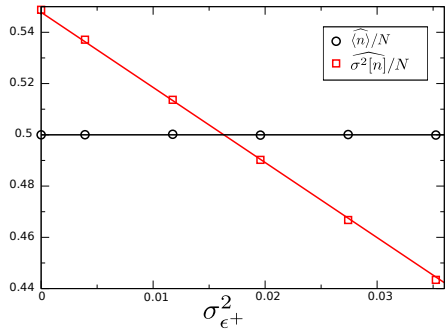}
\caption{Variance and average of the number of agents in state $1$ as a function of the variance of the spontaneous transition rate to state $1$, $\epsilon^+$, in Kirman's model. Results coming from numerical simulations (symbols) and theoretical analysis (solid lines, Eqs.(\ref{medkirmanprefer}, \ref{varkirmanprefer})), for $N=50$ agents, $\lambda_i=\lambda=0.5$ and $\epsilon_++\epsilon_-=2\epsilon=0.4$. $\epsilon_i^+$ are independent random variables distributed according to a symmetric beta distribution in the interval $(0,\epsilon)$ with mean $\overline{\epsilon^+}=0.2$ and variance, $\sigma^2_{\epsilon^+}$.
 \label{fig:kirmanpreference}} 
\end{figure}

Interestingly, other ways to introduce heterogeneity in the system have different effects:\newline
-If the heterogeneity is introduced in the spontaneous transition rate, $\epsilon\to\epsilon_i$, making some particles more prone to spontaneous transitions that others  (but keeping $\lambda_j=\lambda$, $\forall j$ to isolate effects), collective fluctuations again increase with respect to the case of identical particles. \newline
-Next, we can assume that the rate of induced change is different for different agents, even if all have the same influence. Measuring this difference in ``susceptibility" (to induced change) with a parameter $\omega_i$, we would have that the rate of induced change in agent $i$ is $\omega_i\sum_j\lambda_j(1-\delta_{s_i,s_j})/N$.
The effect of heterogeneity in $\omega_i$ (keeping again $\lambda_j=\lambda$, $\forall j$) is that the collective fluctuations decrease with the degree of heterogeneity in the susceptibility $\omega_i$.\newline
-Setting some heterogeneous preference for the states among the particles, i.e. making $\epsilon_i^{+}$, the spontaneous rate from $0$ to $1$ of particle $i$, different from $\epsilon_i^{-}$, the spontaneous rate from $1$ to $0$ of the same particle, decreases global fluctuations. In order to vary the preference for one state keeping constant the global ``intrinsic noise" of this particle (note that the correlation time of particle $i$, when isolated, is given by $\epsilon_i^++\epsilon_i^-$), we set $\epsilon_i^{+}=2\epsilon-\epsilon_i^-$ and generate $\epsilon_i^-$ as  i.i.d. random variables with a distribution with support contained in the interval $[0,2\epsilon]$. Exact explicit expressions for the first moments of the global variables are (see Supplementary Information):\newline
\begin{eqnarray}
 \langle n\rangle_{st}&=&N\frac{\overline{\epsilon^+}}{2\epsilon}\label{medkirmanprefer}\\
\sigma^2[n]_{st}&=&\frac{N}{4(\epsilon+\frac{2\lambda}{N})}\left[\overline{\epsilon^+}(1+\frac{\lambda}{\epsilon})-\frac{\overline{\epsilon^+}^2}{\epsilon}\left(\frac{\lambda}{2\epsilon}+1\right)-2\frac{\sigma^{2}[\epsilon]}{2\epsilon+\lambda}\right]\label{varkirmanprefer},
\end{eqnarray}
In figure (\ref{fig:kirmanpreference}) the exact expressions (\ref{medkirmanprefer}, \ref{varkirmanprefer}) are compared with results coming from numerical simulations.

In this case, the correlation function decays exponentially,
\begin{equation}
 K[n](t)=\sigma^2[n]_{\textrm{st}}e^{-2\epsilon t},
\end{equation}
independently of the degree of heterogeneity, so this form of heterogeneity cannot be inferred by measuring the correlation function. Numerical simulations confirm this result.

\subsubsection{Intuitive explanation of the effects of heterogeneity}

We have seen that heterogeneity can have an ambivalent effect over the size of the fluctuations, depending on the particular form it appears. We now provide intuitive arguments to understand these different effects.

When the influence parameter, $\lambda_i$, varies from one unit to the other, there will be some largely influential agents and others with little influence. In the limit of very large heterogeneity we can think of a situation with a single agent with an extremely large influence and the others having a negligible one (we are keeping a constant average influence). In this case, the highly influential agent drifts from one state to the other, essentially independently (since other agents have negligible influence), but, due to its large influence, all the agents are attracted to its current state. In this ``follow the leader" regime, we obtain macroscopic transitions from one state to the other, corresponding to very large global fluctuations. 

The situation is the opposite for a non-identical susceptibility parameter $\omega_i$ where global fluctuations decrease as the diversity is increased. Again, we can understand this in the limit of very large heterogeneity where a single agent (or a small number of them) has large susceptibility while all the others have a negligible one (in order to keep average susceptibility constant). Then, agents with small susceptibility change essentially independently, in an uncorrelated fashion, resulting in low global fluctuations (note that in order to have large global fluctuations, the fluctuations in the state of the single agents should be correlated).

In the case of diverse spontaneous transition rates, $\epsilon_i$, global fluctuations increase with the degree of heterogeneity. In the limit of large heterogeneity, we would have a small number of agents with very large spontaneous transition rate, whose state would fluctuate in an uncorrelated fashion, and a large number of agents with low spontaneous transition rate, that essentially would only change state through induced transitions, giving rise to correlated fluctuations, resulting in large variance for the global variable.

In the case in which agents display an intrinsic heterogeneous preference for one of the two states, the global fluctuations decrease with heterogeneity degree. We saw this already in the first section for non-interacting agents. Here we see the same effect, suggesting that the phenomenon is robust and still plays a role when interaction is added.
 
The coexistence of a small number of agents with large value for the parameter and a large number with a small value assumed in the previous arguments arises from the fact that an unbounded (from above) distribution of positive-defined parameters (e.g. rates) is skewed.  However, all the effects of diversity commented are still present if the distribution is symmetric. In this case, nevertheless, the maximum degree of heterogeneity (for a constant mean value) is bounded, sometimes greatly limiting the maximum possible value of diversity. For symmetric distributions, a simple explanation is not so clear, but an asymmetry in the effect of increasing and decreasing the value of the parameter seems to be at the heart of the phenomenon.   

\subsection{Application to the SIS disease spreading model}
The previous example could be treated exactly because, due to symmetry, the interaction, non-linear terms, cancel out in the equations for the moments. In general, however, this is not the case, and the analytical treatment is more involved. Here we consider an example of such case.
The stochastic susceptible-infected-susceptible (SIS) model is a paradigmatic model for the study of spreading of infectious disease \cite{SIS} as well as the diffusion of  innovation \cite{pyoung} and other types of social influence. Despite its simplicity, it captures interesting phenomenology. The process is schematically described by:
\begin{equation}
S(i)+I(j) {{ \lambda_j/N \atop \longrightarrow}\atop{}} I(i)+I(j), I(j) {{\gamma \atop \longrightarrow}\atop{}} S(j), S(j) {{\epsilon \atop \longrightarrow}\atop{}} I(j),
\end{equation}
where $S(i)$ (resp. $I(i))$ denotes agent $i$ being susceptible (resp. infected). There are 3 basic elementary processes: (i) infected agent $j$ infects susceptible agent $i$ at a rate $\lambda_j/N$, being $\lambda_j$ the infectivity parameter of agent $j$; (ii) infected agent $j$ becomes susceptible a rate $\gamma$; (iii) susceptible agent $j$ gets infected spontaneously (due to interactions with agents not considered in the system or other causes) at a rate $\epsilon$. This corresponds to the SIS model with spontaneous contagions and distributed infectivity. In the absence of spontaneous infections $\epsilon=0$, the system has a trivial steady state with zero infected agents. With $\epsilon\neq0$ the system has a non-trivial steady state whose properties we analyze in the following. As in the previous case, heterogeneity could appear in any parameter of the agents (for example, in the recovery rate, in a``susceptibility'' parameter, etc.). 

\begin{figure}
\centering
\includegraphics[scale=0.25,angle=0,clip]{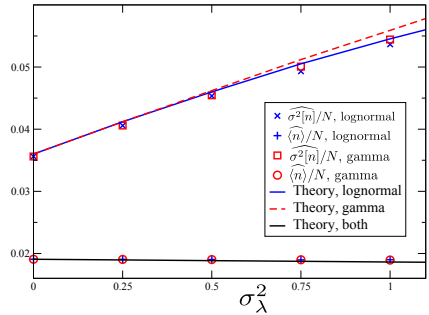}
\caption{Average and variance of the number of infected agents in the SIS model as a function of the variance of the infectivity. Numerical simulations (symbols) and theoretical prediction to first order (lines). Parameters values are $\epsilon=0.01$, $\gamma=1$, $N=200$, $\overline{\lambda}=0.5$.% $\lambda_i$ are i.i.d. random variables with average value $\overline{\lambda}=0.5$ and variance, $\sigma^2[\lambda]$, indicated in the figure.  
 \label{sismed}} 
\end{figure}

We study first the case in which only the infectivity, $\lambda_i$, can vary from agent to agent. The effect of heterogeneity in the deterministic version of related models was studied recently \cite{epidemic}. 
The master equation is of the form (\ref{mastereqgen}) with rates $r_i^+=\epsilon+\sum_l\frac{\lambda_l\langle s_l\rangle}{N}, r_i^-=\gamma$. Equations (\ref{momentsgeneral}-\ref{correlationsgeneral}) for the first moments
can be closed in the steady state, using our main ansatz, to obtain explicit formulas for $\langle n\rangle_\textrm{st}$ and $\sigma^2[n]_\textrm{st}$ to any desired order in $N^{-1}$. In this case, however, the expressions are rather cumbersome and we skip them here. The results are plotted in figure (\ref{sismed}), where we compare the approximation to order $O(N^{-1})$ with results coming from numerical simulations, showing good agreement. Here both the average value and the variance are modified by the presence of heterogeneity (the dependence of the average is, however, only in second order in $1/N$, almost unnoticeable in the figure). As in the Kirman model, the size of the fluctuations increase markedly with the amount of heterogeneity in the ``influence'' (now influence to infecting others) of the agents.

In this case, other ways to introduce heterogeneity also have different effects. When heterogeneity appears in the recovery rate $\gamma$, the mean number of infected agent increases, with a moderate effect over the variance (resulting in smaller relative fluctuations). Heterogeneity in the susceptibility to infection (which would be introduced with the change $r_i^+=\epsilon+\sum_l\frac{\lambda_l\langle s_l\rangle}{N}\rightarrow\epsilon+\omega_i\sum_l\frac{\lambda_l\langle s_l\rangle}{N}$, with $\omega_i$ distributed over the population) decreases the fluctuations, with little effect over the mean value. Heterogeneity in the spontaneous infection rate $\epsilon$ has almost no effect.
In a real situation, one expects to find heterogeneity simultaneously in several of the parameters defining the model. When heterogeneity is present both in the infectivity and in the susceptibility, the effects of both types of heterogeneity essentially add up, with the size of the fluctuations increasing with the heterogeneity in the infectivity for a given level of heterogeneity in the susceptibility and fluctuations decreasing with the level of heterogeneity in the susceptibility for a given level of heterogeneity in the infectivity. 
The effects of heterogeneity in the infectivity and in the susceptibility are equivalent to those found in the Kirman model, and can be intuitively understood in the same terms. Heterogeneity in the recovery rate is similar to assigning an heterogeneous preference for the state $0$ (recovery) and its effect in the (relative) fluctuations is again the same as that in the case of the Kirman model. This suggests that the effects of the heterogeneity found are generic and can be useful to understand the behavior of other systems.
\section{Discussion}
In this work, we have analyzed the combined effect of stochasticity and heterogeneity in interacting-particle systems. We have presented a formulation of the problem in terms of master equations for the individual units, but extracted conclusions about the fluctuations of collective variables. We have developed an approximation suitable for the analytical study of this general type of systems. We have shown that the heterogeneity can have an ambivalent effect on the fluctuations, enhancing or decreasing them depending on the form of the system and the way heterogeneity is introduced. In the case of independent particles, heterogeneity in the parameters always decreases the size of the global fluctuations. We have also demonstrated that it is possible to obtain precise information about the degree and the form of the heterogeneity present in the system by measuring only global variables and their fluctuations, provided that the underlying dynamical equations are known. 
In this way stochastic modeling allows to obtain information not accessible from a purely deterministic approach. 
We have also demonstrated that, in some cases, one can account for the heterogeneity of the particles without losing analytical tractability.

Heterogeneity among the constituent units of a system is a very generic feature, present in many different contexts and this work provides a framework for the systematic study of the effect of heterogeneity in stochastic systems, having thus a wide range of potential applicability. More research in this direction would be welcomed.
\section{Methods}
We have developed and used analytical tools based on an extension of van Kampen's ansatz on the relative weight of the fluctuations compared to the mean value, suitable for systems with particle heterogeneity. We have included the details of this method in the Supplementary Information. In some cases, and in order to compare with the analytical expressions, we have generated data from numerical simulations using a particular form of Gillespie's algorithm \cite{gillespie} that takes into account the heterogeneity in the population. We now explain this algorithm using, for the sake of concreteness, the specific case of the Kirman model with distributed susceptibility.

The parameters of the system are: the spontaneous transition rate $\epsilon$, the influence parameter $\lambda$, the susceptibility parameter of each agent $\omega_i$, and the total number of agents $N$. In this case, the influence parameter can be reabsorbed rescaling $\omega_i$, so we set $\lambda=1$ without loss of generality. The variables of the system are the state of each agent $s_i=0,1$. We will also use the total number of agents in state $1$, $n=\sum s_i$, the total susceptibility of agents in state $1$, $\Omega=\sum s_i\omega_i$, and the average susceptibility $\overline{\omega}=\sum\frac{\omega_i}{N}$.

At any given instant, two events can happen:\newline
(i) An agent in state $1$ changes to state $0$. This can happen due to a spontaneous transition, at a total rate $n\epsilon$, or due to an induced transition, at a total rate $\Omega\frac{(N-n)}{N}$.\newline
(i) An agent in state $0$ changes to state $1$. This can happen due to a spontaneous transition, at a total rate $(N-n)\epsilon$, or due to an induced transition, at a total rate $(N\overline{\omega}-\Omega)\frac{n}{N}$.

According to the Gillespie method, that considers the continuous-time process, the time at which the next transition will take place is exponentially distributed, with average the inverse of the total rate. The probability that a given transition is realized is proportional to its rate. If the realized transition is a spontaneous one, the agent that actually undergoes it is selected at random (since, in this case, they all have the same rate $\epsilon$). If the transition is induced, the agent that undergoes it is selected with probability proportional to its susceptibility. It can be easily seen that this principles lead to an exact (up to numerical precision) simulation of sample paths of the stochastic process \cite{gillespie}.

The algorithm, then, proceeds as follows:\\
(0) Evaluate the total number of particles in state $1$, $n$, and the total susceptibility of particles in state $1$, $\Omega$.\newline
(1) Evaluate the total transition rate $r=\epsilon N+\Omega\frac{N-n}{N}+(N\overline{\omega}-\Omega)\frac{n}{N}$.\newline
(2) Generate the time for the next reaction, $t_n$, as an exponential random variable with average $1/r$. This can be done by setting $t_n=-\frac{1}{r}\ln U$, with $U$ a uniform random variable in the range $(0, 1)$.\newline
(3) Select which reaction takes place. For this, generate a uniform random variable, $g$, in the range $(0,r)$. \\
\indent-If $g<n\epsilon,$ the transition will be a spontaneous transition from $1$ to $0$; Select an agent, $j$, at random among those at state $1$. Set $n=n-1$, $\Omega=\Omega-\omega_j$.\newline
\indent-If $n\epsilon\leq g<N\epsilon$, the transition will be a spontaneous transition form $0$ to $1$; Select an agent, $j$, at random among those at state $0$. Set $n=n+1$, $\Omega=\Omega+\omega_j$.\newline
\indent-If $N\epsilon\leq g<N\epsilon+\frac{\Omega(N-n)}{N}$, the transition will be an induced transition from $1$ to $0$; Select an agent, $j$, among those at state $1$ with probability proportional to the value of its susceptibility parameter $\omega_j$. Set $n=n-1$, $\Omega=\Omega-\omega_j$.\newline
\indent-If $ g\ge N\epsilon+\frac{\Omega(N-n)}{N}$, the transition an induced transition from $0$ to $1$; Select an agent, $j$, among those at state $0$ with probability proportional to the value of its susceptibility parameter $\omega_j$. Set $n=n+1$, $\Omega=\Omega+\omega_j$.\newline
(4) Set $t=t+t_n$. Go to (1). 

\begin{acknowledgments}
We thank E. Hernandez-Garcia for useful discussions. This work was supported by MINECO (Spain), Comunitat Aut\`onoma de les Illes
Balears, FEDER, and the European Commission under project FIS2007-60327. L.F.L. is supported by the JAEPredoc program of CSIC.
\end{acknowledgments}

\newpage
{\centering {\Large \textbf{Supplementary Information}}}

\subsection{M-states system}
We consider here the case in which each particle can be in one of $M$ (instead of $2$) possible states. We will show that the results obtained in the main text for $2-$state systems also hold in this more general case. 

We label the states with the subscript $\alpha=0,1,\dots,M-1$, so in this case the variable describing the state of particle $i$ can take $M$ possible values, $s_i=0,\dots,M-1$ (we start the labeling from $0$ to be consistent with the previous case, that would correspond to $M=2$).
Let $p_i(\lambda_i,\alpha,t)$ the probability that particle $i$, with heterogeneity parameter $\lambda_i$, be on state $\alpha$. It satisfies the evolution equation:
\begin{equation}
 \frac{d p_i(\lambda_i,\alpha,t)}{dt}=\sum_\beta A_{\alpha,\beta}(\lambda_i)p_i(\lambda_i,\beta,t)\label{multiplestates},
\end{equation}
with $A_{\alpha,\beta}$ a general transition matrix (satisfying $\sum_{\gamma=0}^{N-1}A_{\gamma,\alpha}=0$), that may depend in principle on time and on the time that the particle has been on its current state.
To isolate the role of parameter heterogeneity, we assume that the initial condition is the same for all the particles (or that the initial condition is determined by the value of $\lambda_i$) such that the solution $p_i(\lambda_i,\alpha,t)=p(\lambda_i,\alpha,t)$ is the same for all particles sharing the same value of the parameter.
The macroscopic state of the system will be described by the set of variables $n_\alpha=\sum_{i=1}^{N}\delta_{\alpha,s_i}$, that is, the number of particles in each state. The averages and variances of this variables are given by:
\begin{eqnarray}
 \langle n_\alpha(t)\rangle&=&\sum_{i=1}^N p(\lambda_i,\alpha,t)\\
\sigma^2[n_\alpha(t)]&=&\sum_{i=1}^N\left[ p(\lambda_i,\alpha,t)-p(\lambda_i,\alpha,t)^2\right].
\end{eqnarray}
This variance is again smaller that tat of a system of identical particles with same average, the difference given by:
\begin{equation}
 \sigma^2[n_\alpha(t)]_\text{id}-\sigma^2[n_\alpha(t)]=N\overline{p(\alpha,t)^2}-\overline{p(\alpha,t)}^2,
\end{equation}
a result exactly analogous to the one obtained in the previous case. The heterogeneity among the particles on the probability of occupation of level $\alpha$ can be derived from the first moments of the occupation number of the level:
\begin{equation}
 \overline{p(\alpha,t)^2}-\overline{p(\alpha,t)}^2=\frac{\langle n_\alpha\rangle-\langle n_\alpha\rangle^2/N-\sigma^2[n_\alpha]}{N}.
\end{equation}
Note that, when focusing on the number of particles on state $\alpha$, the system effectively reduces to a $2-$level one, with states $\alpha$ and no-$\alpha$, so the results of the previous section can be translated directly. 

A different and some times relevant question can be considered when the labeling of the states is such that the order is well defined (for example each state corresponds to an energy level or a distance from a reference). Then the average state is meaningful and we can study its statistical properties. Below we show that the variance of this mean level is again always smaller if heterogeneity is present.

The average state of the system is given by $L=\sum_{\alpha=0}^{M-1}\alpha\ \displaystyle \frac{n_\alpha}{N}$. It is a random variable whose average and variance are given by:
\begin{eqnarray}
 \langle L\rangle&=&=\sum_{\alpha=0}^{M-1}\alpha\frac{\langle n_\alpha\rangle}{N}=\sum_{\alpha=0}^{M-1}\sum_{i=1}^{N}\alpha\frac{p(\lambda_i,\alpha)}{N},\\
\sigma^2[L]&=&\sum_{\alpha,\beta=0}^{M-1}\frac{\alpha\beta}{N^2}(\langle n_\alpha n_\beta\rangle-\langle n_\alpha\rangle\langle n_\beta\rangle)=\frac{1}{N^2}\sum_{i=1}^N\left[\sum_{\alpha=0}^{M-1}\alpha^2p(\alpha,\lambda_i)-\sum_{\alpha,\beta=0}^{M-1}\alpha p(\alpha,\lambda_i)\beta p(\beta,\lambda_i)\right]\label{meanlevelhet}.
\end{eqnarray}
We have used $p(\lambda_i,\alpha)=\langle\delta_{\alpha,s_i}\rangle$ and $\langle n_\alpha n_\beta\rangle=\sum_{i,j=1}^N\langle\delta_{\alpha,s_i}\delta_{\beta,s_j}\rangle=\langle n_\alpha\rangle\langle n_\beta\rangle+\sum_{i=1}^N[\delta_{\alpha,\beta}p(\alpha,\lambda_i)-p(\alpha,\lambda_i)p(\beta,\lambda_i)]$.
A system of identical particles that had the same average occupation of the different levels i.e. $p_{id}(\lambda_i,\alpha)=\frac{1}{N}\sum_{j=1}^Np(\lambda_j,\alpha,)=\frac{\langle n_\alpha\rangle}{N}$ $\forall i,\alpha$, would have and average and variance of the mean level given by:
\begin{eqnarray}
 \langle L\rangle_{id}&=&\sum_{\alpha=0}^{M-1}\alpha\frac{\langle n_\alpha\rangle}{N}=\langle L\rangle,\\
\sigma^2[L]_\text{id}&=&\frac{1}{N}\sum_{\alpha=0}^{M-1}\alpha^2\frac{\langle n_\alpha\rangle}{N}-\frac{1}{N}\sum_{\alpha,\beta=0}^{M-1}\alpha\beta\frac{\langle n_\alpha\rangle}{N}\frac{\langle n_\beta\rangle}{N}\label{meanlevelid}.
\end{eqnarray}
We now define $g(\lambda_i)\equiv\sum_\alpha\alpha p(\lambda_i,\alpha)$ (the average level of particle $i$), and note that the first terms in the right-hand side of (\ref{meanlevelhet}) and (\ref{meanlevelid}) are equal, while the second terms can be written as:
\begin{eqnarray}
\frac{1}{N^2}\sum_{i=1}^N\sum_{\alpha,\beta=0}^{M-1}\alpha p(\lambda_i,\alpha)\beta p(\lambda_i,\beta)&=&\frac{1}{N^2}\sum_{i=1}^Ng(\lambda_i)^2=\frac{1}{N}\overline{g^2},\\
\frac{1}{N}\sum_{\alpha,\beta=0}^{M-1}\alpha\beta\frac{\langle n_\alpha\rangle}{N}\frac{\langle n_\beta\rangle}{N}&=&\frac{1}{N}\left[\frac{1}{N}\sum_{i=1}^Ng(\lambda_i)\right]^2=\frac{1}{N}\overline{g}^2,
\end{eqnarray}
which implies that $\sigma^2[L]_\text{id}\ge \sigma^2[L]$, i.e. the variance of the mean level is always smaller in a system of heterogeneous particles, the difference with respect to the case of identical ones being:
\begin{equation}
 \sigma^2[L]_\text{id}-\sigma^2[L]=\frac{1}{N}\left(\overline{g^2}-\overline{g}^2\right)=\frac{1}{N}\sum_{\alpha,\beta=0}^{M-1}\alpha\beta\left[\sum_{i=1}^N\frac{p(\alpha,\lambda_i)p(\beta,\lambda_i)}{N}-\sum_{i,j=1}^N\frac{p(\alpha,\lambda_i)p(\beta,\lambda_j)}{N^2}\right]\geq0.
\end{equation}
The correction to the variance in this case scales as $1/N$, but again is of the same order as the variance itself, indicating a non-negligible correction.
In this case to derive the heterogeneity of $g(\lambda_i)$ over the population one needs to know the average occupation level of each state $\langle n_\alpha\rangle$ and use:
\begin{equation}
 \overline{g^2}-\overline{g}^2=\sum_\alpha\alpha^2\langle n_\alpha\rangle/N-\langle L\rangle^2-N\sigma^2[L].
\end{equation}
This can be written in terms of the variance of $L$ in an equivalent system of identical particles, $\sigma^2[L]_\text{id}$. If this is known, one can directly use
\begin{equation}
 \overline{g^2}-\overline{g}^2=N\left(\sigma^2[L]_\text{id}-\sigma^2[L]\right).
\end{equation}
Note that, contrary to the two-level case, now the value of $\langle L\rangle$ does not determine $\sigma^2[L]_\text{id}$.

\subsection{Intuitive origin of the decrease of fluctuations for independent units}
We have shown that a system of independent heterogeneous particles has smaller fluctuations for the collective variable than an equivalent system of identical ones. The origin of this result is the following (for simplicity we refer to the case of $2$-state system):

The average of the global variable is determined by the concentration of the states of the particles around state $1$ ($\langle n\rangle=\sum_i\langle s_i\rangle$). The fluctuations (measured by the variance) of the global variable are determined by the stochastic fluctuations of the individual particles alone ($\sigma^2[n]=\sum_i\sigma^2[s_i]$, since the particles are independent).\newline
In a system of heterogeneous particles, the dispersion of the states of the particles is due to the heterogeneity (some prefer to be around sate $0$, others prefer to be around sate $1$) plus their intrinsic stochasticity. 
In a system of identical particles, the dispersion comes from the stochasticity alone, so for a system of identical particles to have the same concentration in the states of the particles (global average) than a heterogeneous system, the intrinsic stochasticity has to be larger. This will give rise to larger fluctuations for the global variable.\newline
In particular, any given rational value of $\frac{\langle n\rangle}{N}=\frac{A}{B}$ can be obtained with zero fluctuations, taking $A$ particles that are always at state $1$ and $B-A$ particles that are always at state $0$.

\begin{figure}[h]
  \centering
  \includegraphics[scale=0.3,angle=0,clip]{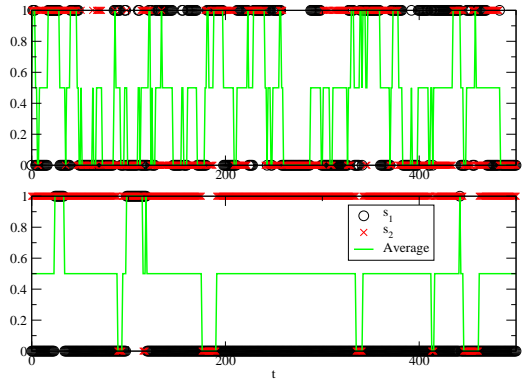}
  \caption{Time series of a system of two identical (upper panel) and heterogeneous (lower panel) particles, together with the corresponding average state. Note that the fluctuations of the average state are more pronounced in the case of the identical particles. 
  \label{fig:trajectories}} 
  \end{figure}

This explanation is illustrated in figure (\ref{fig:trajectories}). In the identical-particles system both particles fluctuate between $1$ and $0$. In the heterogeneous case, one particle spends most of the time at $1$ and the other spends most of the time at $0$. The probability of finding a given particle at $1$ is the same in both cases ($1/2$) but in the heterogeneous case most of the time there is one particle at $1$ and one particle at $0$, resulting on a value of the average state most often equal to $1/2$, and so with smaller fluctuations.
The situation is similar for a larger number of particles, as  shown in figure 1 of the main text. % (\ref{fig:trajectoriesindep}).
An analogous picture emerges when one considers more that $2$ states. Note that in every case we compare a system of heterogeneous particles with another of identical ones that has the same one-particle distribution i.e. $p_i(\alpha)_{id}=\sum_j\frac{p(\alpha,\lambda_j)}{N}$, $\forall i,\alpha$.
% \begin{figure}[h]
%   \centering
%   \includegraphics[scale=0.4,angle=0,clip]{trajecindep}
%   \caption{Time series the global variable of a system of identical (left panels) and heterogeneous (right panels) particles. The upper panels correspond to $N=10$ particles and the lower ones to $N=100$ particles. The parameters $p_i$ of the particles have been chosen from a symmetric Beta distribution with average equal to $\overline{p}=0.5$ and variance indicated in the figure (zero in the case of identical particles), this was done by setting $r^+_i=1, r^-_i=t⁺(1/p_i-1)$. Note that the fluctuations of the average state are more pronounced in the case of the identical particles. 
%   \label{fig:trajectoriesindep}} 
%   \end{figure}

\subsection{Justification of the Ansatz}
The general evolution equations for the first moments are of the form:
\begin{eqnarray}
 \frac{d\langle s_i\rangle}{dt}&=&\langle r_i^{+}\rangle-\langle (r_i^{-}+r_i^{+})s_i\rangle,\label{momentsgeneral}\\
\frac{d\langle s_is_j\rangle}{dt}&=&-\langle q_{ij}s_js_i\rangle+\langle r_i^{+}s_j\rangle+\langle r_j^{+}s_i\rangle.\label{correlationsgeneral}
\end{eqnarray}
Our main ansatz is that the $m$-particle correlations $\sigma_{j_1,...,j_m}(t)=\langle \delta_{j_1}(t)\cdots\delta_{j_m}(t)\rangle$ with $\delta_j(t)= s_j(t)-\langle n_{j}(t)\rangle$
 scale with system size as 
\begin{equation}
 \sigma_{j_1,...,j_m}(t)=O(N^{-m/2}),\hspace{0.3cm} \text{for } j_k\neq j_l.\label{ansatz}
\end{equation}
We first show how the ansatz (\ref{ansatz}) allows to close the system (\ref{momentsgeneral}, \ref{correlationsgeneral}).\newline
We assume that functional dependence of the rates on the sate variables is of the form $f(s_1/N,\dots,s_N/N)$. This includes, for example, rates of the form $f(\sum \lambda_ks_k/N)$ like the ones used in the examples analyzed. We further assume that the rates can be expanded as a power series:
\begin{equation}
 f(s_1/N,\dots,s_N/N)=a_0+\sum_{i_1=1}^N a_{i_1}\frac{s_{i_1}}{N}+\frac{1}{2!}\sum_{i_1,i_2=1}^{N} a_{i_1,i_2}\frac{s_{i_1}s_{i_2}}{N^2}+\dots+\frac{1}{k!}\sum_{i_1,\dots,i_k=1}^{N} a_{i_1,\dots,i_k}\frac{s_{i_1}\cdots s_{i_k}}{N^k}+\dots
\end{equation}
There are $N^k$ terms in the $k$'th summand, $\displaystyle\sum_{i_1,\dots,i_k=1}^N$, giving a total contribution of order $O(N^0)$. 
%As this is divided by the factor $1/k!$, it is possible to truncate the series at some finite $k$.\newline
%the leading dependence of the rates are on the scaled variables $s_i/N$, as it is usually the case as we will see in the explicit examples below (a common factor of any order in $N$ multiplying all the rates is acceptable, since it can be absorbed rescaling the time). 
The terms in the right hand side of (\ref{momentsgeneral}) are of the form:
\begin{equation}
 \frac{\langle s_{i_1}\dots s_{i_k}\rangle}{k!}=\frac{\langle(\delta_{i_1}+\langle s_{i_1}\rangle)\dots(\delta_{i_k}+\langle s_{i_k}\rangle)\rangle}{k!}=\sum_{l=0}^k
\frac{\delta^l\langle s\rangle^{k-l}}{l!(k-l)!}=\sum_{l=0}^k\frac{O(N^{-l/2})}{l!(k-l)!},
\end{equation}
where $\delta^l$ corresponds to a term of the form $\langle \delta_{j_1}(t)\cdots\delta_{j_l}(t)\rangle$, $\langle s\rangle^{k-l}$ corresponds to $\langle s_{i_1}\rangle\cdots\langle s_{i_{k-l}}\rangle$ and the last equality holds due to our ansatz. We see that the dominant terms are those with $l=0$, which correspond to products of mean values of the form $\langle s_{i_1}\rangle\cdots\langle s_{i_k}\rangle$. We conclude that the ansatz allows to do the substitution $\langle s_{i_1}\dots s_{i_k}\rangle\rightarrow\langle s_{i_1}\rangle\cdots\langle s_{i_k}\rangle+O(N^{-1/2})$ in the evolution equations for the mean values.

The evolution equations for the correlations read:
\begin{equation}
 \frac{d\sigma_{i,j}}{dt}=\langle(r^-_i+r^+_i)s_i\delta_j\rangle+\langle(r^-_j+r^+_j)s_j\delta_i\rangle+\langle r^+_i\delta_j\rangle+\langle r^+_i\delta_j\rangle.\label{generalcorrelations}
\end{equation}
In this case, the terms are of the form $\langle s_{i_1}\dots s_{i_k}\delta_{r}\rangle=\langle(\delta_{i_1}+\langle s_{i_1}\rangle)\dots(\delta_{i_k}+\langle s_{i_k}\rangle)\delta_{r}\rangle$ with $r=i,j$. Due to the presence of $\delta_{s}$, the term in which only averages appears vanishes. Reasoning as before, we see that 
the dominant terms are those proportional to $\sigma_{i_l,s}$, while those proportional to higher-order correlations can be neglected. In this case, the ansatz allows to do the substitution $\langle s_{i_1}\dots s_{i_k}\delta_{r}\rangle\rightarrow\langle s_{i_1}\rangle\cdots\langle s_{i_k}\rangle\displaystyle\sum_{l=1}^k\frac{\sigma_{i_r}}{\langle s_{i_r}\rangle}+O(N^{-3/2})$.
In this way, the evolution equation for the correlations depend, at first order, only on averages and correlations and not on higher order moments.

The validity of the ansatz (\ref{ansatz}) itself can be established a posteriori by checking that the results obtained using the ansatz are consistent with it.
In this section, we will link its validity with the well-known van Kampen's ansatz \cite{VK} that is the basis for the systematic system-size expansion.

Van Kampen's ansatz consists on assuming that the variable of interest has a deterministic part of order $\Omega$ plus a stochastic part of order $\Omega^{1/2}$, i.e. $n=\Omega\phi(t)+\Omega^{1/2}\xi$, where $\Omega$ is a parameter of the system that controls the relative size of the changes due to elementary processes, typically the system size.

In our system the role of the parameter $\Omega$ is played by the total number of particles $N$. As briefly stated in the main text, we cannot expect that the single-particle variables that we are considering obey van Kampen's ansatz, since they are not extensive. Our variables $s_i=0,1$ have a deterministic and stochastic part that are both of order zero respect to $N$ (note that $\sigma^2[s_i]=\langle s_i\rangle(1-\langle s_i\rangle)$). However, the macroscopic variable $n=\sum s_i$ is indeed extensive and we can expect that it will follow van Kampen's ansatz: $n=N\phi(t)+N^{1/2}\xi$. This implies that the $m$-th central moment of $n$ will scale as $N^{m/2}$, i.e:
\begin{equation}
 \langle(n-\langle n\rangle)^m\rangle=\sum_{j_1,...,j_m}\sigma_{j_1,...,j_m}=O(N^{m/2}).\label{vkansatzcorr}
\end{equation}
Now, assuming that $\sigma_{j_1,...,j_m}=f_{m}(N)\tilde{\sigma}_{j_1,...,j_m}$ for $j_k\neq j_l$, with $\tilde{\sigma}_{j_1,...,j_m}$ independent of $N$ i.e. the $m$-particle correlations are all or the same order in $N$, so that $\sum_{j_1\neq j_2\neq,...,\neq j_m}\tilde{\sigma}_{j_1,...,j_m}$ scales as $N^m$ (note that there are of the order of $N^{m}$ terms in the sum), we obtain our main ansatz, $\sigma_{j_1,...,j_m}=O(N^{-m/2})$ for $j_k\neq j_l$. We have only considered terms with $j_k\neq j_l$ in the sum (\ref{vkansatzcorr}); terms with repeated sub-indexes can be expressed as lower order ones. For example, if the index $j_1$ is present $k$ times, and the others are all different, we find:
\begin{eqnarray}
 &&\sigma_{j_1,j_1,...,j_1,j_2,...j_{m-k+1}}=\langle(s_{j_1}-\langle s_{j_1}\rangle)^k\delta_{j_2}\dots\delta_{j_{k-k+1}}\rangle\nonumber\\
&&=\sigma_{j_2,...j_{m-k+1}}(-\langle s_{j_1}\rangle)^k+\langle\delta_{j_2}\dots\delta_{j_{m-k+1}}\sum_{i=0}^{k-1}\binom{k}{i}(-\langle s_{j_1}\rangle)^is_{j_1}\rangle\\
&&=\sigma_{j_2,...j_{m-k+1}}[(1-\langle s_{j_1}\rangle)^k\langle s_{j_1}\rangle+(1-\langle s_{j_1}\rangle)(-\langle s_{j_1}\rangle)^k]+\sigma_{j_1,\dots,j_{m-k+1}}[(1-\langle s_{j_1}\rangle)^k-(-\langle s_{j_1}\rangle)^k]\nonumber
\end{eqnarray}
as can be see expanding $(s_{j_1}-\langle s_{j_1}\rangle)^k$ and keeping in mind that $s_i^2=s_i$. The number of such terms in the sum (\ref{vkansatzcorr}) is $O(N^{m-k+1})$, so they give smaller contribution that terms with all sub-indexes different. Proceeding order by order from $k=1$, we see that our main ansatz (\ref{ansatz}) follows from (\ref{vkansatzcorr}). 
 \newline
We point out that in systems of heterogeneous particles we do not have a closed description for the global, extensive, variable $n$ so van Kampen's expansion cannot be used. Instead we derive the implications of van Kampen's ansatz over the correlations of the microscopic variables. (\ref{ansatz}) is a simple and convenient expression that in general allows to close the equation for the moments (\ref{momentsgeneral},\ref{correlationsgeneral}). Often, however it is not necessary, and a weaker condition of the form (\ref{vkansatzcorr}), that directly follows from van Kampen's ansatz without further assumptions, is sufficient.

Van Kampen's ansatz is generally valid when the macroscopic equations have a single attracting fixed point, when the system displays small fluctuations around the macroscopic state. The general method explained here is expected to be valid under similar conditions. An interesting topic for future research will be whether a system that has a single attracting fixed point in the absence of diversity always maintains this globally stable state when diversity is present, and whether a system that does not posses this globally stable fixed point can acquire it when diversity is added.

\subsection{Details of the calculation for the Kirman model}
In the Kirman model with distributed influence, the averages and correlations obey:
\begin{eqnarray}
 \frac{d\langle s_i\rangle}{dt}&=&\epsilon-(2\epsilon+\overline{\lambda})\langle s_i\rangle+N^{-1}\sum_{k}\lambda_k\langle s_k\rangle\label{Kirmanaverage},\\
\frac{d\sigma_{i,j}}{dt}&=&-2(2\epsilon+\overline{\lambda})\sigma_{i,j}+N^{-1}\sum_k\lambda_k\left(\sigma_{i,k}+\sigma_{j,k}\right)\nonumber\\
&+&\delta_{i,j}\left[\epsilon+a+(\overline{\lambda}-2a)\langle s_i
\rangle-2\sum_k\frac{\lambda_k\sigma_{i,k}}{N}\right]\label{Kirmancorrelations}
\end{eqnarray}
with $a\equiv\sum_k\frac{\lambda_k\langle n_k\rangle}{N}$. Note that, due to the particular form of the rates, these equations are indeed closed. The first equation leads to a steady state value $\langle s_i\rangle_\text{st}=\frac{1}{2}$, which implies $\langle n\rangle_\text{st}=\frac{N}{2}$ (a property that comes from the symmetry $0\leftrightarrow1$). (\ref{Kirmancorrelations}) is a linear system of equations for the correlations. The steady state correlations can always be obtained by inverting the matrix that gives the couplings. Obtaining a closed expression for $\sigma^2[n]$ in terms of the moments of $\lambda$ is, however, not straightforward.
From (\ref{Kirmancorrelations}), we see that in the steady state:
\begin{equation}
 \sigma_{i,j}=\frac{\sum_k\lambda_k\frac{\sigma_{i,k}+\sigma_{j,k}}{N}+\delta_{i,j}\left[\epsilon+\overline{\lambda}/2-2\sum_k\frac{\lambda_k\sigma_{i,k}}{N}\right]}{2(2\epsilon+\overline{\lambda})}\label{sigmakir},
\end{equation}
from where we obtain
\begin{equation}
\sigma^2[n]=\sum_{i,j}\sigma_{i,j}=\frac{N(\epsilon+\overline{\lambda}/2)+2C(1-1/N)}{2(2\epsilon+\overline{\lambda})}\label{sigmakir2},
\end{equation}
with $C\equiv\sum_{i,j}\lambda_j\sigma_{i,j}$. Multiplying (\ref{sigmakir}) by $\lambda_j$ and summing over $j$, one obtains:
\begin{eqnarray}
 C&=&\frac{d(1-2/N)+(\epsilon+\overline{\lambda}/2)\overline{\lambda} N}{4\epsilon+\overline{\lambda}},\label{C}
 %e&=&\frac{\overline{\lambda^2}d+(\epsilon+\overline{\lambda}/2)N\overline{\lambda^3}N-2\sum_{i,k}\frac{\lambda_i^3\lambda_k\sigma_{i,k}}{N}}{4\epsilon+\overline{\lambda}}\label{e},
\end{eqnarray}
where $d\equiv\sum_{i,j}\lambda_i\lambda_j\sigma_{i,j}$. This is obtained again multiplying (\ref{sigmakir}) by $\lambda_i\lambda_j$ and summing over $i,j$: 
\begin{eqnarray}
 d&=&\frac{(\epsilon+\overline{\lambda}/2)\langle\lambda^2\rangle N-2e/N}{4\epsilon}\label{d},
 %e&=&\frac{\overline{\lambda^2}d+(\epsilon+\overline{\lambda}/2)N\overline{\lambda^3}N-2\sum_{i,k}\frac{\lambda_i^3\lambda_k\sigma_{i,k}}{N}}{4\epsilon+\overline{\lambda}}\label{e},
\end{eqnarray}
where $e\equiv\sum_{i,j}\lambda_i^2\lambda_j\sigma_{i,j}$. Using the ansatz $\sigma_{i,j}=O(N^{-1})$ we see that the last term of (\ref{d}) is $O(N^0)$ (while the other are of $O(N)$), so to the first order we obtain :
\begin{equation}
 \sigma^2_\text{st}[n]=\frac{N}{4}\left[1+\frac{\overline{\lambda}}{2\epsilon}+\frac{\sigma^2[\lambda]}{2\epsilon\left(4\epsilon+\overline{\lambda}\right)}\right]+O(N^0),\label{kirman1st}
\end{equation}
with $\sigma^2[\lambda]=\overline {\lambda^2}-\overline\lambda^2$.
% \begin{equation}
%  \sigma^2=\frac{N}{4}\left(1+2\frac{\frac{\langle\lambda^2\rangle}{4\epsilon}+\overline{\lambda}}{4\epsilon+\overline{\lambda}}\right)+O(N^0)
% \end{equation}
% It is possible to calculate this correction to any desired order in $1/N$, introducing more terms. Higher order corrections depend on higher moments of the distribution of parameters. Next one gives:
% \begin{equation}
%  \sigma^2=N\frac{\langle\lambda^2\rangle+6\epsilon\overline{\lambda}+8\epsilon^2}{8\epsilon(\overline{\lambda}+4\epsilon)}+\frac{\langle\lambda^2\rangle^2+2\langle\lambda^2\rangle(3\overline{\lambda}-4\epsilon)\epsilon-4\epsilon(\langle\lambda^3\rangle+2\overline{\lambda}\epsilon(\overline{\lambda}+4\epsilon))}{16\epsilon^2(\lambda_s+4\epsilon)^2}+O(N^{-1})
% \end{equation}
We have seen how the application of the ansatz (\ref{ansatz}) allows one to obtain closed expression for the global average and variance. Interestingly, in this particular example, it is possible to include all higher order terms to obtain an exact expression for $d$ (which gives the exact expression for $\sigma^2[n]$ trough (\ref{C},\ref{sigmakir2})), details are given in the appendix:
\begin{equation}
  d=\frac{N(\epsilon+\overline{\lambda}/2)\sum_{k=0}^{\infty}\left(\frac{-2}{N(4\epsilon+\overline{\lambda})}\right)^k\overline{\lambda^{2+k}}}{4\epsilon+\overline{\lambda}-\sum_{k=0}^{\infty}\left(\frac{-2}{N(4\epsilon+\overline{\lambda})}\right)^k\overline{\lambda^{1+k}}}=\frac{N(\epsilon+\overline{\lambda}/2)\overline{\frac{\lambda^2}{1+\frac{2\lambda}{N(4\epsilon+\overline{\lambda})}}}}{4\epsilon+\overline{\frac{2\lambda^2}{N(4\epsilon+\overline{\lambda})+2\lambda}}}\label{dexact}
 \end{equation}
The second equality holds as long as $\lim_{m\rightarrow\infty}\overline{\frac{\lambda^{m+2}}{1+\frac{2\lambda}{N(4\epsilon+\overline{\lambda})}}}\left(\frac{2}{N(4\epsilon+\overline{\lambda})}\right)^m=0$ (note that $\displaystyle\sum_{k=0}^Ma^k\overline{\lambda^{k+2}}=\overline{\lambda^2\displaystyle\sum_{k=0}^M a^k\lambda^k}=\overline{\lambda^2\displaystyle\frac{1-a^{k+1}\lambda^{k+1}}{1-a\lambda}}$). A sufficient condition for this is $\lambda_i<\frac{(\overline{\lambda}+4\epsilon)N}{2},\ \forall i=1,\dots,N$. When the $\lambda_i$'s are i.i.d. random variables, the probability that this condition is satisfied approaches one as $N$ grows. This condition is actually necessary and sufficient for the first equality in (\ref{dexact}) to hold (see appendix).\newline
We finally obtain the following exact expression for the variance:
\begin{equation}
 \sigma^2_\text{st}[n]=\frac{N}{4}\left[1 + \frac{2\overline{\lambda}(1-1/N)}{4\epsilon+\overline{\lambda}} + (N-3+2/N)\frac{\overline{\frac{\lambda^2}{N(4\epsilon+\overline{\lambda})+2\lambda}}}{2\epsilon+\overline{\frac{\lambda^2}{N(4\epsilon+\overline{\lambda})+2\lambda}}}\right]\label{kirmanexact}
\end{equation}

We see from (\ref{kirmanexact}) that higher order corrections to $\sigma^2[n]$ depend on higher order moments of the distribution of $\lambda$ over the population. 

Expressions (\ref{kirman1st}, \ref{kirmanexact}) refer to the variance of $n$ in a population with given values for the parameters of each agent, $\lambda_i$, so the averages are population averages i.e. $\overline{g(\lambda)}=\sum_{i=1}^Ng(\lambda_i)/N$. In the case that the parameters of the agents are random variables, the population averages themselves, $\overline{g(\lambda)}$, become random variables. To compute the expected (average) value of (\ref{kirman1st}, \ref{kirmanexact}), $\widehat{\sigma^2[n]}$, one has to average over the distribution of $\overline{g(\lambda)}$, which depends on the distribution $f(\lambda)$ of the $\lambda_i's$ (we are assuming $\lambda_i's$ i.i.d. random variables). This averages were obtained numerically, by evaluating expressions (\ref{kirman1st}, \ref{kirmanexact}) over the same realizations of the $\lambda_i$'s that were used in the numerical simulations. One can use the approximation $\widehat{\overline{g(\lambda)}}\simeq\widehat{g(\lambda)}$, that works better the larger the $N$ and the lower the variance $\sigma^2_{\lambda}$, and that, due to the law of large numbers, is valid in the limit $N\rightarrow\infty$. In Fig.2  of the main text we compare the average of the analytical expression (\ref{kirmanexact}) with results coming from numerical simulations. We find perfect agreement and see that at first order the dependence of $\sigma^2[n]$ with $\sigma^2_\lambda\equiv\widehat{\lambda^2}-\widehat{\lambda}^2$ is linear and independent of the form of the distribution, as indicated by (\ref{kirman1st}). Higher order corrections are noticeable for higher levels of diversity.

In the case of heterogeneity in the preference of the agents for the states, as indicated in the main text, the variance is given by:
\begin{eqnarray}
 \sigma^2[n]_{st}&=&\frac{N}{4(\epsilon+\frac{2\lambda}{N})}\left[\overline{\epsilon^+}(1+\frac{\lambda}{\epsilon})-\frac{\overline{\epsilon^+}^2}{\epsilon}\left(\frac{\lambda}{2\epsilon}+1\right)-2\frac{\sigma^{2}[\epsilon]}{2\epsilon+\lambda}\right]\label{varkirmanprefer},
\end{eqnarray}
In this case, the average of (\ref{varkirmanprefer}) over the distribution of parameters can be easily computed, giving:
\begin{eqnarray}
\widehat{\sigma^2[n]}_{st}&=&\frac{N}{4(\epsilon+\frac{2\lambda}{N})}\Bigg[\widehat{\epsilon^+}\left(2+\frac{\lambda}{\epsilon}\right)-\widehat{\epsilon^+}^2\left(\frac{\lambda}{2\epsilon^2}+\frac{1}{\epsilon}\right)\nonumber\\
&&\hspace{1.5cm}-\sigma^2_{\epsilon^+}\left(\frac{2\epsilon+\lambda/N}{\epsilon(2\epsilon+\lambda)}+\frac{\lambda}{2\epsilon^2N}\right)\Bigg]\label{varkirmanprefermed},
\end{eqnarray}
\newline

The correlation function can be derived as follows (we exemplify the derivation in the case of distributed influence, for other types of heterogeneity, the derivation is similar):\newline
(\ref{Kirmanaverage}) is an equation for the conditional averages $\langle s_i|\{s_l(t_0)\}\rangle$ if we set $\{s_l(t_0)\}$ as initial conditions. It implies:
\begin{equation}
 \frac{da}{dt}=\epsilon\lambda-2\epsilon a\rightarrow a(t_0+t)=\frac{\lambda}{2}(1-e^{-2\epsilon t})+a(t_0)e^{-2\epsilon t},
\end{equation}
with $a\equiv\sum_k\lambda_k\langle s_k|\{s_l(t_0)\}\rangle/N$.                      
Noticing that (\ref{Kirmanaverage}) is equal to $\frac{d\langle s_i\rangle}{dt}=\epsilon-(2\epsilon+\lambda)\langle s_i\rangle+a(t)$, we obtain:
\begin{equation}
 \langle s_i(t_0+t)|\{s_k(t_0)\}\rangle=\frac{1}{2}(1-e^{-(2\epsilon+\overline{\lambda})t})+\frac{a(t_0)-\overline{\lambda}/2}{\overline{\lambda}}e^{-2\epsilon t}(1-e^{-\overline{\lambda}t})+s_i(t_0)e^{-(2\epsilon+\overline{\lambda})t}.
\end{equation}
Using now $K_{st}[n](t)=\langle \langle n(t_0+t)|n(t_0)\rangle n(t_0)\rangle_{st}-\langle n\rangle_{st}^2=\sum_{i,j}\langle\langle s_i(t_0+t)|\{s_k(t_0)\}\rangle s_j(t_0)\rangle-\frac{N^2}{4}$ (remember $\langle n\rangle_{st}=N/2$), and after some straightforward algebra, we obtain:
\begin{equation}
 K_{st}[n](t)=(\sigma_{st}^2-C/\overline{\lambda})e^{-(2\epsilon+\overline{\lambda})t}+C/\overline{\lambda}e^{-2\epsilon t}.\label{kst}
\end{equation}
From (\ref{sigmakir}) we get $C/\overline{\lambda}=\frac{2\epsilon+\overline{\lambda}}{2\overline{\lambda}(1-1/N)}(\sigma_{st}^2-N/4)\equiv u$, showing that (\ref{kst}) is equal to the expression displayed in the main text.
\subsection{Appendix}\label{kirmanappendix}
We start with equation (\ref{sigmakir}):
\begin{equation}
 \sigma_{i,j}=\frac{\sum_k\lambda_k\frac{\sigma_{i,k}+\sigma_{j,k}}{N}+\delta_{i,j}\left[\epsilon+\overline{\lambda}/2-2\sum_k\frac{\lambda_k\sigma_{i,k}}{N}\right]}{2(2\epsilon+\overline{\lambda})}.
\end{equation}
Using the rescaled variables $\tilde{\sigma}_{i,j}\equiv4\sigma_{i,j}, \tilde{\lambda}_k\equiv\frac{\lambda_k}{2(2\epsilon+\overline{\lambda})N}$, and defining $S_n:=\sum_{i,j=0}^N\tilde{\lambda_i}^n\tilde{\lambda}_j\tilde{\sigma}_{i,j}$, we obtain:
\begin{equation}
S_{n+1}=\frac{N\overline{\tilde{\lambda}}-1}{2}S_n+\frac{N}{2}\left(\overline{\tilde{\lambda}^n}S_1+\overline{\tilde{\lambda}^{n+1}}\right).
\end{equation}
Defining now $G_n:=\left(\frac{2}{N\overline{\tilde{\lambda}}-1}\right)^nS_n, T_M:=\sum_{n=1}^MG_n$, we arrive to:
\begin{eqnarray}
 G_{n+1}&=&G_n+\left(\frac{2}{N\overline{\tilde{\lambda}}-1}\right)^{n+1}\frac{N}{2}\left[G_1\left(-\frac{\overline{\lambda}+4\epsilon}{4(2\epsilon+\overline{\lambda})}\right)\overline{\tilde{\lambda}^{n}}+\overline{\tilde{\lambda}^{n+1}}\right],\\
 T_{M+1}-G_1&=&T_M+\frac{N}{2}\sum_{n=1}^M\left[\left(\frac{2}{N\overline{\tilde{\lambda}}-1}\right)^n\left(\frac{2\overline{\tilde{\lambda}^{n+1}}}{N\overline{\tilde{\lambda}}-1}+G_1\overline{\tilde{\lambda}^{n}}\right)\right].
\end{eqnarray}

If $\lim_{M\rightarrow\infty}G_M=0$, we see that:
\begin{equation}
 G_1=-\frac{\frac{N}{2}\sum_{n=1}^{\infty}\left(\frac{2}{N\overline{\tilde{\lambda}}-1}\right)^{n+1}\overline{\tilde{\lambda}^{n+1}}}{1+\frac{N}{2}\sum_{n=1}^{\infty}\left(\frac{2}{N\overline{\tilde{\lambda}}-1}\right)^{n}\overline{\tilde{\lambda}^{n}}}.
\end{equation}
Going back to the original variables, we finally obtain, with the notation of the main text:
\begin{equation}
 d=\frac{\frac{N^3(\epsilon+\overline{\lambda}/2)(4\epsilon+
\overline{\lambda})}{4}\sum_{n=1}^{\infty}\left(\frac{-2}{(\overline{\lambda}+4\epsilon)N}\right)^n\overline{\lambda^{n+1}}}{1+\frac{N}{2}\sum_{n=1}^{\infty}\left(\frac{-2}{(\overline{\lambda}+4\epsilon)N}\right)^n\overline{\lambda^n}},
\end{equation}
which can be rewritten in the form (\ref{dexact}), completing the proof.

The condition of convergence is:
\begin{equation}
 \lim_{M\rightarrow\infty}G_M=\lim_{M\rightarrow\infty}\sum_{i,j=1}^N\left(\frac{-2\lambda_i}{(\overline{\lambda}+4\epsilon)N}\right)^M\frac{2\lambda_j}{(2\epsilon+\overline{\lambda})N}\sigma_{i,j}=0.
\end{equation}
A necessary and sufficient condition for this is $\lambda_i<\frac{(\overline{\lambda}+4\epsilon)N}{2}, \forall i=1,\dots N$. When the parameters $\lambda_i$ are i.i.d. r. v. the probability of this typically approaches $1$ as $N$ grows. 


\begin{thebibliography}{}
\bibitem{diversitycells} Spudich, J. L., Koshland, Jr. D. E. Non-genetic individuality: chance in the single cell. {\it Nature} \textbf{262,} 467-471 (1976).
\bibitem{diversitycellsrev} Snijder, B., Pelkmans, L. Origins of regulated cell-to-cell variability. {\it Nat. Rev. Mol. Cell Biol} \textbf{12,} 119-125 (2011).
\bibitem{granovetter} Granovetter, M. Threshold Models of Collective Behavior. {\it Am. J. Sociol.} \textbf{83,} 1420-1443 (1978).
\bibitem{kirmanrepagent} Kirman, A. Whom or What Does the Representative Individual Represent? {\it J. Econ. Perspect} \textbf{6,} 117-136 (1992).
\bibitem{nonidlasers} Braiman, Y., Kennedy, T. A. B. , Wiesenfeld, K. and Khibnik, A.  Entrainment of solid-state laser arrays. {\it Phys. Rev. A} \textbf{52,} 1500-1506 (1995). 
\bibitem{nonidlasersstrogatz} Oliva, R. A. and Strogatz, S. H.  Dynamics of a large array of globally coupled lasers with distributed frequencies. {\it Int. J. Bifurcat. Chaos} \textbf{11,} 2359-2374 (2001).
\bibitem{barabasirev}Albert, R. and Barab\'asi, A.-L. Statistical mechanics of complex networks. {\it Rev. Mod. Phys.} \textbf{74,} 47-97 (2002).
\bibitem{yamirnetworks} Boccaletti, S., Latora, V., Moreno, Y., Chavez, M., Hwang, D.-U. Complex networks: Structure and dynamics {\it Phys. Rep.} \textbf{424,} 175-308 (2006).
\bibitem{divresonance} Tessone, C. J., Mirasso, C. R., Toral. R. and Gunton, J. D. Diversity-induced resonance. {\it Phys. Rev. Lett.} \textbf{97,} 194101 1-4 (2006).
\bibitem{quenchednoise} Komin, N., Lacasa, L. and Toral, R. Critical behavior of a Ginzburg–Landau model with additive quenched noise. {\it J. Stat. Mech.} \textbf{P12008,} 1-19 (2010).
\bibitem{pyoung} Peyton Young, H. Innovation Diffusion in Heterogeneous Populations: Contagion, Social Influence, and Social Learning. {\it Am. Econ. Rev.} \textbf{99,} 1899-1924 (2009).
\bibitem{epidemic} Novozhilov, A. S. Epidemiological Models With Parametric Heterogeneity: Deterministic Theory for Closed Populations {\it Math Model Nat Pheno} \textbf{7,} 147-167 (2012).
\bibitem{redner} Masuda, N., Gibert, N. and Redner, S. Heterogeneous voter models. {\it Phys. Rev. E} \textbf{82,} (010103R) 1-4 (2010).
\bibitem{mobilia} Mobilia, M. and Georgiev, I. T., Voting and catalytic processes with inhomogeneities. {\it Phys. Rev. E} \textbf{71,} 046102 1-17 (2005).
\bibitem{minor} Xie, J. and Sreenivasan, S. Social consensus through the influence of committed minorities. {\it Phys. Rev. E} \textbf{84,} 011130 1-8 (2011).
%\bibitem{minorities} Xie, J., Sreenivasan, S., Korniss, G., Zhang, W., Lim, C. and Szymanski1, B. K.. Social consensus through the influence of committed minorities. {\it Phys. Rev. E} \textbf{84, }011130 1-8 (2011).
\bibitem{fisher} Fisher, D. S. Critical-behavior of random transverse-field Ising sping chains {\it Phys. Rev. B} \textbf{51,} 6411-6461 (1995).
\bibitem{Young} Young, A. P. (Ed.) {\it Spin Glasses and Random Fields} World Scientific, Singapore (1998).
\bibitem{Parisi} M\'ezard, M., Parisi, G. and Virasoro, M. A. {\it Spin Glass Theory and Beyond} World Scientific, Singapore (1987).
\bibitem{bouchaud90} Bouchaud, J-.P. and Georges, A., Anomalous diffusion in disordered media: statistical mechanisms, models and physical applications. {\it Phys. Rep.} \textbf{195,} 127-293 (1990).
\bibitem{avraham-havling00} Ben-Avraham, D. and Havlin, S. {\it Diffusion and Reactions in Fractals and Disordered Systems} Cambridge University Press, Cambridge (2000).
\bibitem{parametric_heterogeneity} Dushoff, J. Host Heterogeneity and Disease Endemicity: A Moment-Based Approach. {\ Theor Popul Biol} \textbf{56,} 325–335 (1999).
\bibitem{VK} van Kampen, N. G. \textit{Stochastic Processes in Physics and Chemistry}, North-Holland, Amsterdam (2004).
\bibitem{kirman} Kirman, A. Ants, Rationality, and Recruitment. {\it Q. J. Econ.} \textbf{108,} 137-156 (1993). 
\bibitem{votermodel} Liggett, T. M. \textit{Interacting Particle Systems}, ͑Springer-Verlag, New York (1985).
\bibitem{alfarano} Alfarano, S. and Milakovi\'c, M. Network structure and N-dependence in agent-based herding models. {\it J. Econ. Dynam. Control} \textbf{33,} 78-92 (2009).
\bibitem{SIS} Anderson, R. M. \textit{Population Dynamics of Infectious Diseases: Theory and Applications}, Chapman and Hall, London-New York (1982).
\bibitem{gillespie} Gillespie, D. T. \textit{Exact stochastic simulation of coupled chemical reactions} \textbf{81,}(25) 2340-2361 (1977).
\end{thebibliography}
\end{document}